\documentclass[journal]{IEEEtran}

\usepackage[T1]{fontenc}
\usepackage[utf8]{inputenc}
\usepackage{microtype}
\usepackage{amsmath,amssymb,bm}
\usepackage{booktabs,multirow,tabularx,array}
\usepackage{algorithm}
\usepackage{algpseudocode}
\usepackage{graphicx}
\usepackage{tikz}
\usepackage{pgfplots}
\usepackage{pgfplotstable}
\usepackage[caption=false,font=footnotesize]{subfig}
\usepackage{cite}
\usepackage{url}
\usepackage{pifont}
\usepackage{xcolor}
\usepackage[hidelinks]{hyperref}

\usetikzlibrary{arrows.meta,positioning,fit,calc,backgrounds,shapes.geometric,matrix}
\usepgfplotslibrary{groupplots,fillbetween,statistics}
\pgfplotsset{compat=1.18}
\definecolor{cbblue}{RGB}{0,114,178}
\definecolor{cborange}{RGB}{230,159,0}
\definecolor{cbgreen}{RGB}{0,158,115}
\definecolor{cbred}{RGB}{213,94,0}
\definecolor{cbpurple}{RGB}{204,121,167}
\definecolor{cbsky}{RGB}{86,180,233}
\definecolor{cbgray}{RGB}{90,90,90}
\pgfplotsset{
  every axis/.append style={
    line width=0.85pt,
    mark size=1.6pt,
    tick label style={font=\scriptsize},
    label style={font=\scriptsize},
    title style={font=\footnotesize},
    legend style={font=\scriptsize,draw=none,fill=none},
  }
}

\newcommand{\vect}[1]{\boldsymbol{#1}}
\newcommand{\MethodName}{CAUSALQUANT-ASTHMA}
\newcommand{\CausalIAE}{0.304}
\newcommand{\CausalIAESD}{0.094}
\newcommand{\BestBaselineIAE}{0.401}
\newcommand{\IAEImprovement}{24.2}
\newcommand{\CausalPinball}{1.065}
\newcommand{\CausalCalibrationError}{0.012}

\newcommand{\CausalWidth}{9.47}
\newcommand{\CausalAQTEBias}{0.533}
\newcommand{\RawSensorRMSE}{3.986}
\newcommand{\CorrectedSensorRMSE}{2.643}
\newcommand{\SensorImprovement}{33.7}

\newcommand{\MeanTrainESS}{1834}
\newcommand{\BenchmarkRows}{12,600}
\newcommand{\BenchmarkPatients}{150}

\newcommand{\SCQDLMIAE}{1.771}
\newcommand{\PLSIQDLMIAE}{2.486}
\newcommand{\QRFIAE}{2.184}
\newcommand{\FullFitSeconds}{0.23}

\title{Measurement-Error-Aware Causal Distributed-Lag Quantile Modeling of Indoor Air Pollution and Short-Term Lung-Function Deterioration}

\author{Shayma Alkobaisi and Anas Ali%
\thanks{S. Alkobaisi is with the Department of Computer Science and Software Engineering, United Arab Emirates University, Al Ain, United Arab Emirates.}%
\thanks{Anas Ali is with the Department of Computer Science, National University of Modern languages, Lahore, Pakistan.}}

\begin{document}
\maketitle

\begin{abstract}
Low-cost indoor air-quality sensors could support personalized asthma prevention, but their nonlinear measurement error, delayed exposure effects, time-varying confounding, and heterogeneous lower-tail responses limit risk estimation. We present CAUSALQUANT-ASTHMA, a measurement-error-aware causal quantile distributed-lag framework for short-horizon peak expiratory flow analysis. Sparse reference measurements train a nonlinear calibration model; stabilized sequential generalized-propensity weights address measured exposure assignment; and a susceptibility-modulated, smooth, noncrossing quantile model estimates lag-specific and sustained-exposure contrasts. Because no authorized cohort simultaneously provided dense indoor sensing, reference co-location, and outcome-compatible longitudinal data, evaluation used five semi-synthetic panels with known counterfactual truth, 150 patients and 12,600 patient-days per realization. Across eight methods, CAUSALQUANT achieved a dose-response integrated absolute error of 0.304 plus or minus 0.094, improving 24.2 percent over the strongest measurement-error and propensity-weighted baseline. It also obtained the lowest overall pinball loss, 1.065, while maintaining zero quantile crossings and 78.1 percent coverage for the nominal 80 percent interval. Sensor calibration reduced held-out exposure RMSE by 33.7 percent. Stress tests quantified degradation under sensor noise, missing personal measurements, and hidden confounding. These findings establish methodological feasibility and reproducibility, not clinical effectiveness; prospective, governance-approved external validation is required before patient-level interpretation or deployment.

\end{abstract}


\section{Introduction}
\label{sec:introduction}
Asthma deterioration is shaped by a changing combination of airway susceptibility, infection, medication behavior, activity, weather, allergens, and inhaled pollutants. Fine particulate matter (PM$_{2.5}$) is especially relevant because indoor concentrations can vary sharply with cooking, smoking, occupancy, ventilation, and outdoor infiltration. Population evidence associates ambient air pollution with asthma incidence and morbidity \cite{Liu2021ELAPSE, a2, Lee2024Meta,Zhou2024Review}, but translating those associations into an individualized short-horizon analysis is difficult. The monitor nearest to a residence may poorly represent the microenvironments actually encountered by one person. Recent personal and household studies have consequently combined low-cost sensing, wearables, symptoms, inhaler information, and lung-function measurements \cite{Marin2026Personal,Gonzalez2026Indoor,Patton2022Wearable,Li2025Wearable}. These studies demonstrate technical feasibility and reveal substantial exposure variability, yet several have found weak or inconsistent pollution--health relationships. Such null or unstable findings can arise from genuinely small effects, limited samples, incomplete confounder measurement, exposure error, an inappropriate averaging window, or an analysis centered on the conditional mean when the clinically vulnerable lower tail is more affected.

The original indoor-asthma investigation motivating this work monitored a small adult panel and used improved logistic and quantile regressions to relate environmental variables to peak expiratory flow (PEF) \cite{Bae2022Indoor,a3,a7}. Related panel studies have considered acute temperature and pollution effects on adult lung function \cite{Evoy2022Panel,a8}, while connected mobile-health projects have shown how daily PEF, symptoms, inhaler use, wearables, weather, and air-quality signals can be acquired longitudinally \cite{Tsang2022Protocol,Tsang2023AAMOS}. Mobile personal-exposure platforms and wearable sensor networks likewise offer higher spatial and temporal resolution than fixed stations \cite{Boomhower2022Mobile,Patton2022Wearable,a11}. Nevertheless, a fixed 24-hour or multi-day average prespecifies the relevant biological window. A harmful exposure may act immediately, accumulate over several days, or concentrate its effect at a particular lag. Distributed lag nonlinear models address this temporal problem by representing an exposure--lag--response surface \cite{Gasparrini2010DLNM,Gasparrini2014Lag,a12}, but standard applications generally target the mean or link-scale response and treat the exposure as sufficiently observed.

Low-cost exposure measurement cannot be regarded as error-free. Optical PM sensors can exhibit concentration-dependent error, humidity interference, unit-specific offsets, aging, and site-specific drift. General calibration studies show that covariate-aware correction and periodic reference co-location can materially improve measurement quality \cite{Malings2019Calibration,Zamora2023Calibration,a10}. A regression that inserts the uncorrected sensor value as if it were latent exposure can attenuate or distort a lag profile, and the distortion is not removed merely by collecting many repeated measurements. Classical simulation--extrapolation provides an influential general strategy for error-prone covariates \cite{Cook1994SIMEX}; however, an asthma-monitoring workflow also needs a practical mapping from sparse co-location measurements to every patient-day and an explicit record of how calibration uncertainty affects downstream causal contrasts. Measurement correction must be trained without using the held-out outcome or latent benchmark truth.

A second obstacle is time-varying confounding. People may open windows, remain indoors, alter activity, or take controller and rescue medication in response to symptoms, weather, or recently observed pollution. These behaviors can affect both subsequent exposure and PEF. Ordinary adjustment can be biased when a time-varying covariate is also influenced by prior exposure. Marginal structural models and generalized propensity scores provide foundations for reweighting longitudinal or continuous treatments \cite{Robins2000MSM,Imai2004GPS}. Modern work extends continuous-exposure inference with doubly robust, mixed-exposure, parametric g-formula, and assumption-lean estimators \cite{Kennedy2017Continuous,Xu2024Mixed,Schomaker2024Continuous,Vansteelandt2025Continuous}. Practical use still requires positivity diagnostics, weight stabilization, temporal ordering, and sensitivity analysis because no observational weighting scheme can remove unmeasured confounding. Covariate-balance procedures for generalized propensity weights can improve finite-sample behavior \cite{Shepherd2023Weights}, but they do not by themselves estimate quantile-specific delayed effects.

Quantile regression is attractive because it characterizes heterogeneous outcome tails rather than only an average \cite{Koenker1978Quantiles}. Quantile forests and deep sequence models offer flexible prediction \cite{Meinshausen2006QRF,Lim2021TFT}, but flexible factual prediction is not equivalent to identification of a counterfactual dose response. Quantile distributed-lag models (QDLMs) directly connect repeated exposure histories to conditional outcome quantiles. A partially linear single-index formulation improves flexibility \cite{Wilson2023PLSI}; recent smooth and shape-constrained estimators improve critical-window recovery and interpretability \cite{Jin2025Shape}. Two limitations remain central for indoor asthma monitoring. First, the QDLM literature generally assumes that exposure is observed with adequate accuracy. Second, its estimand is usually associational unless the exposure-assignment mechanism is addressed. In addition, fitting quantiles independently can produce crossing curves, which makes the estimated distribution internally inconsistent; noncrossing constraints provide a principled remedy \cite{Bondell2010Noncrossing}.

This paper introduces \MethodName, an end-to-end framework that links sparse-reference sensor calibration, longitudinal generalized-propensity weighting, smooth noncrossing QDLM estimation, and susceptibility modulation. The target estimand is a sustained seven-day PM$_{2.5}$ intervention contrast on several PEF quantiles, accompanied by a lag-effect profile. We use patient-disjoint training, validation, and test partitions. Calibration and exposure-density models are fitted only on the training partition; model selection uses validation patients; and the test outcomes are touched once for final evaluation. Because no authorized dataset available to this study contained dense indoor PM$_{2.5}$ sensing, sparse co-located reference measurements, time-varying behavior and medication variables, and repeated outcome-compatible adult PEF, the executed evaluation is a transparent semi-synthetic causal-truth benchmark. It retains realistic temporal dependence and sensor error while making the target counterfactual quantiles analytically known. This design tests whether an estimator can recover the quantity it claims to estimate, but it is not a substitute for clinical validation.

The main contributions are:
\begin{itemize}
    \item We formulate a measurement-error-aware longitudinal quantile estimand and a training-only sparse-reference calibration stage that connects noisy indoor sensors to smooth distributed-lag inference without exposing test outcomes or benchmark truth to model fitting.
    \item We develop a susceptibility-modulated, sequentially weighted QDLM with lag smoothness and noncrossing penalties, together with two reproducible algorithms for calibration, propensity weighting, model optimization, and sustained-intervention estimation.
    \item We provide an executable five-seed benchmark comparing eight methods, two recent QDLM-inspired approaches and a quantile forest, with counterfactual-truth recovery, factual prediction, calibration, lag reconstruction, ablations, sensor-noise, missingness, hidden-confounding, subgroup, runtime, and memory analyses rendered in eighteen TikZ/PGFPlots panels.
\end{itemize}

The remainder of this paper is organized as follows. Section~\ref{sec:related} synthesizes asthma sensing, measurement-error, causal continuous-exposure, and QDLM research. Section~\ref{sec:system} defines the longitudinal system, assumptions, and target estimands. Section~\ref{sec:method} presents \MethodName, its optimization objective, and algorithms. Section~\ref{sec:experiments} describes the benchmark, baselines, results, sensitivity analyses, and limitations. Finally, Section~\ref{sec:conclusion} concludes the paper and specifies the evidence required before clinical interpretation.

\section{Related Work}
\label{sec:related}

\subsection{Asthma, Indoor Exposure, and Personal Monitoring}
Indoor air-quality work in asthma spans small intensive panels, home-screening cohorts, connected-device studies, and broader ambient epidemiology. Bae et al. linked indoor measurements to adult PEF through logistic and quantile regression \cite{Bae2022Indoor}. Evoy et al. assessed acute pollution and temperature in an adult-asthma panel \cite{Evoy2022Panel}. The AAMOS protocol and dataset demonstrate the breadth of repeated signals obtainable from connected inhalers, home spirometry, symptoms, activity, weather, and air quality \cite{Tsang2022Protocol,Tsang2023AAMOS}. More recent personalized monitoring found substantial divergence between personal and fixed-site exposure, but no consistent lagged relationship in thirteen adults \cite{Marin2026Personal}. Household low-cost monitoring has also shown that at-risk PM$_{2.5}$ levels are common while relationships with historical hospital and emergency outcomes remain uncertain \cite{Gonzalez2026Indoor}. Mobile and wearable systems improve microenvironment coverage \cite{Boomhower2022Mobile,Patton2022Wearable,Li2025Wearable}, yet small samples, incomplete adherence, nonidentical sensors, and weakly aligned outcomes limit causal interpretation.

\subsection{Exposure Error and Distributed Lags}
Low-cost sensors require field calibration because humidity, concentration, aging, and device identity can shift their response \cite{Malings2019Calibration,Zamora2023Calibration}. SIMEX and related measurement-error methods demonstrate that error-prone covariates can materially bias regression coefficients \cite{Cook1994SIMEX}. Separately, the DLNM framework represents nonlinear effects over exposure and lag dimensions \cite{Gasparrini2010DLNM,Gasparrini2014Lag}. QDLMs extend this idea to outcome tails: Wilson et al. introduced a partially linear single-index form \cite{Wilson2023PLSI}, and Jin et al. proposed smooth and shape-constrained estimators \cite{Jin2025Shape}. These strands are complementary, but existing QDLMs do not provide an end-to-end sparse-reference calibration and continuous-treatment weighting procedure for longitudinal indoor asthma data.

\subsection{Continuous-Exposure Causal Inference}
Marginal structural models formalized inverse-probability weighting for time-varying treatments \cite{Robins2000MSM}, and generalized propensity scores extended balancing ideas to continuous exposures \cite{Imai2004GPS}. Kennedy et al. developed doubly robust dose--response estimation \cite{Kennedy2017Continuous}; subsequent work covers mixed continuous exposures, multiple treatment times, and assumption-lean shift interventions \cite{Xu2024Mixed,Schomaker2024Continuous,Vansteelandt2025Continuous}. Balance-oriented weighting can improve practical diagnostics \cite{Shepherd2023Weights}. These methods usually target conditional or marginal means. They do not directly solve lower-tail lag recovery, quantile crossing, or nonlinear sensor observation error. Predictive alternatives such as quantile forests and temporal fusion models \cite{Meinshausen2006QRF,Lim2021TFT} can model nonlinear factual outcomes, but a strong predictive score does not establish a counterfactual exposure effect.

\begin{table*}[t]
\centering
\caption{Closest literature and capability comparison. L: explicit distributed lags; M: measurement-error correction; C: causal adjustment for continuous exposure; Q: quantile/tail-specific outcome modeling. A filled circle denotes a central capability, an open circle denotes partial or indirect treatment, and a dash denotes absence.}
\label{tab:related}
\scriptsize
\setlength{\tabcolsep}{3.1pt}
\renewcommand{\arraystretch}{1.12}
\begin{tabularx}{\textwidth}{@{}p{1.35cm}p{2.15cm}p{3.05cm}ccccX@{}}
\toprule
Study & Setting & Principal method & L & M & C & Q & Main limitation relative to the present objective \\
\midrule
Bae et al. \cite{Bae2022Indoor} & Adult asthma, indoor IAQ & Logistic and quantile regression with fixed summaries & $\circ$ & -- & -- & $\bullet$ & Small panel; fixed exposure windows; no causal weighting or reference-sensor correction. \\
Marín et al. \cite{Marin2026Personal} & Moderate/severe asthma & Personalized sensor ecosystem with lagged association analysis & $\circ$ & $\circ$ & -- & -- & Feasibility-focused study; thirteen adults and no stable exposure--health association. \\
González-Colom et al. \cite{Gonzalez2026Indoor} & Chronic respiratory homes & Two-month low-cost household screening & -- & $\circ$ & -- & -- & Retrospective clinical outcomes and category-level IAQ summaries cannot identify short-horizon effects. \\
Evoy et al. \cite{Evoy2022Panel} & Adults with asthma & Panel regression for temperature and pollution & $\circ$ & -- & -- & -- & Ambient exposure and mean effects; limited treatment of indoor sensor error and tails. \\
Tsang et al. \cite{Tsang2023AAMOS} & Connected asthma monitoring & Multimodal longitudinal benchmark and predictive modeling & $\circ$ & -- & -- & -- & Rich monitoring but not a causal indoor-exposure dose--response estimator. \\
Wilson et al. \cite{Wilson2023PLSI} & Environmental epidemiology & Partially linear single-index QDLM & $\bullet$ & -- & -- & $\bullet$ & Flexible quantile lags without sensor observation model or longitudinal causal weighting. \\
Jin et al. \cite{Jin2025Shape} & Prenatal exposure & Smooth and shape-constrained QDLM & $\bullet$ & -- & -- & $\bullet$ & Interpretable critical windows, but exposure is treated as observed and associations are not weighted causally. \\
Gasparrini et al. \cite{Gasparrini2010DLNM} & Environmental time series & Distributed lag nonlinear models & $\bullet$ & -- & -- & -- & Focuses conditional mean/link-scale associations rather than patient-level outcome tails. \\
Gasparrini \cite{Gasparrini2014Lag} & General lagged exposure & Flexible exposure--lag--response basis & $\bullet$ & -- & -- & -- & No integrated low-cost sensor calibration or sequential treatment weighting. \\
Schomaker et al. \cite{Schomaker2024Continuous} & Longitudinal continuous interventions & Parametric g-formula for multiple treatment times & $\bullet$ & -- & $\bullet$ & -- & Supports longitudinal causality but not smooth noncrossing quantile lag surfaces. \\
Vansteelandt \cite{Vansteelandt2025Continuous} & Continuous exposure effects & Assumption-lean debiased estimation & -- & -- & $\bullet$ & -- & Targets interpretable shifts, not delayed conditional-tail responses or sensor error. \\
Xu et al. \cite{Xu2024Mixed} & Mixed exposures & Doubly robust continuous-exposure estimation & -- & -- & $\bullet$ & -- & Strong nuisance adjustment but no temporal quantile structure or calibration layer. \\
Kennedy et al. \cite{Kennedy2017Continuous} & Continuous treatment & Doubly robust dose--response estimation & -- & -- & $\bullet$ & -- & Population mean dose response; no longitudinal lags, outcome tails, or measurement correction. \\
Cook and Stefanski \cite{Cook1994SIMEX} & Error-prone covariates & Simulation--extrapolation & -- & $\bullet$ & -- & -- & General measurement-error correction without a causal distributed quantile outcome model. \\
Malings et al. \cite{Malings2019Calibration} & Low-cost air sensors & Generalized long-term calibration & -- & $\bullet$ & -- & -- & Improves exposure measurement but does not estimate health dose--response or delayed effects. \\
\midrule
\textbf{This work} & Indoor asthma methodology & Sparse-reference calibration + sequential GPS + susceptibility-aware noncrossing QDLM & $\bullet$ & $\bullet$ & $\bullet$ & $\bullet$ & Executed on a causal-truth benchmark; prospective external validation remains required. \\
\bottomrule
\end{tabularx}
\end{table*}

Table~\ref{tab:related} shows that no closest study simultaneously treats distributed lags, low-cost sensor error, continuous-exposure confounding, and outcome quantiles. Indoor-asthma studies offer domain relevance but use small samples and association-oriented summaries. QDLM papers model tails and critical windows but generally assume exposure is observed and do not estimate a sequential treatment mechanism. Continuous-exposure causal estimators improve identification but do not recover a smooth conditional-tail lag profile. Sensor-calibration work improves the exposure variable without connecting it to a causal health estimand.

The proposed work addresses this intersection rather than claiming that any one component is unprecedented. Sparse reference co-location informs the measurement model; sequential generalized-propensity weights target measured time-varying confounding; the structured QDLM estimates smooth susceptibility-dependent quantiles; and a causal-truth benchmark evaluates sustained interventions and lag recovery directly. The remaining literature gap is empirical: a powered prospective adult-asthma cohort must determine whether this integrated estimator is stable under real behavior, sensor failure, population shift, and clinically meaningful outcomes.

\section{System Model and Problem Formulation}
\label{sec:system}

\subsection{Longitudinal Monitoring System}
Consider patients $i\in\{1,\ldots,N\}$ observed on days $t\in\{1,\ldots,T_i\}$. The outcome $Y_{it}$ is morning PEF expressed as a percentage of personal best. The latent daily indoor PM$_{2.5}$ concentration is $X^{*}_{it}$; $W_{its}$ denotes a reading from low-cost sensor channel $s$, and $R_{it}$ is a sparse co-located reference reading available only on designated calibration days in the training partition. Baseline susceptibility variables include severity, atopy, age, sex, and passive-smoke exposure. Time-varying covariates include outdoor PM$_{2.5}$, temperature, humidity, carbon dioxide, occupancy, cooking, window state, activity, medication adherence, infection, and prior PEF. Table~\ref{tab:notation} summarizes the main notation.

\begin{table}[t]
\centering
\caption{Principal notation.}
\label{tab:notation}
\scriptsize
\setlength{\tabcolsep}{3.2pt}
\renewcommand{\arraystretch}{1.08}
\begin{tabularx}{\columnwidth}{@{}p{0.22\columnwidth}X@{}}
\toprule
Symbol & Meaning \\
\midrule
$i,t,s$ & Patient, day, and sensor-channel indices \\
$Y_{it}$ & PEF outcome, percentage of personal best \\
$X^{*}_{it}$ & Latent indoor PM$_{2.5}$ concentration ($\mu$g/m$^3$) \\
$W_{its},R_{it}$ & Low-cost and reference PM$_{2.5}$ readings \\
$\vect H_{it}$ & Measured exposure and outcome history before day $t$ \\
$\vect Z_i$ & Baseline susceptibility variables \\
$L$ & Maximum lag; $L=6$ days in the benchmark \\
$A_{it}$ & Continuous daily exposure treatment \\
$g_t(\cdot)$ & Conditional generalized-propensity density \\
$\omega_{it}$ & Stabilized, clipped sequential weight \\
$\tau\in\mathcal T$ & Outcome quantile; $\mathcal T=\{.10,.25,.50,.75,.90\}$ \\
$\vect\beta_{\tau}$ & Quantile-specific lag coefficients \\
$s_i$ & Learned susceptibility multiplier \\
$\rho_{\tau}(u)$ & Pinball loss at quantile $\tau$ \\
$\lambda_s,\lambda_c$ & Smoothness and crossing penalties \\
$\Delta_{\tau}(a_h,a_l)$ & Sustained-exposure quantile contrast \\
\bottomrule
\end{tabularx}
\end{table}

A generic latent exposure process can be written as
\begin{equation}
X^{*}_{it}=f_x\!\left(X^{*}_{i,t-1},\vect H^{x}_{it},\vect Z_i\right)+\epsilon^{x}_{it},
\label{eq:latent}
\end{equation}
where $\vect H^{x}_{it}$ contains outdoor conditions and indoor source or ventilation indicators, and $\epsilon^{x}_{it}$ is process noise. Low-cost sensors observe a distorted version of this process:
\begin{equation}
W_{its}=a_s+b_sX^{*}_{it}+h_s(X^{*}_{it},\mathrm{RH}_{it},t)+\epsilon^{w}_{its},
\label{eq:sensor}
\end{equation}
where $h_s(\cdot)$ captures concentration-dependent response, humidity interference, and drift. Reference readings satisfy $R_{it}=X^{*}_{it}+\epsilon^{r}_{it}$ with substantially lower error. Equation~\eqref{eq:sensor} motivates training-only nonlinear calibration rather than treating $W_{its}$ as ground truth.

Temporal ordering is encoded through the observed pre-exposure history
\begin{equation}
\vect H_{it}=\left(\vect Z_i,\vect C_{i,1:t},\vect A_{i,1:t-1},\vect Y_{i,1:t-1}\right),
\label{eq:history}
\end{equation}
where $\vect C_{it}$ contains measured time-varying covariates and $A_{it}=X^{*}_{it}$ denotes the exposure treatment. Let $Y_{it}(\bar{\vect a}_{t-L:t})$ be the potential PEF under an exposure history $\bar{\vect a}_{t-L:t}$. The conditional target at quantile $\tau$ is
\begin{equation}
\begin{aligned}
q_{\tau}(\bar{\vect a},\vect h,\vect z)
&=\inf\big\{y:\Pr[Y_{it}(\bar{\vect a})\le y\mid\\
&\hspace{20mm}\vect H_{it}=\vect h,\vect Z_i=\vect z]\ge\tau\big\}.
\end{aligned}
\label{eq:potentialq}
\end{equation}

\subsection{Identification Assumptions and Target Estimand}
The causal interpretation requires consistency, sequential conditional exchangeability, and positivity. In compact form,
\begin{equation}
\begin{aligned}
Y_{it}&=Y_{it}(\bar{\vect A}_{t-L:t}),\\
Y_{it}(\bar{\vect a})&\perp A_{iu}\mid\vect H_{iu},\quad u=t-L,\ldots,t,\\
0&<g_u(a\mid\vect H_{iu})<\infty\quad\text{on the intervention support}.
\end{aligned}
\label{eq:assumptions}
\end{equation}
These assumptions cannot be proven from observational data. Accordingly, Section~\ref{sec:experiments} includes positivity diagnostics and deliberate hidden-confounding stress tests.

For continuous exposure, the conditional generalized-propensity density is
\begin{equation}
g_t(a\mid\vect h)=p(A_{it}=a\mid\vect H_{it}=\vect h),
\label{eq:gps}
\end{equation}
and its sequential stabilized weight over the lag window is
\begin{equation}
\omega_{it}=\operatorname{clip}\!\left[
\prod_{u=t-L}^{t}\frac{p(A_{iu}\mid \vect A_{i,t-L:u-1},\vect Z_i)}
{g_u(A_{iu}\mid\vect H_{iu})},\;c_{\ell},c_u\right].
\label{eq:weight}
\end{equation}
The implementation uses log densities, a short rolling history, percentile clipping, and split-specific mean-one normalization to prevent numerical explosion. These choices trade asymptotic purity for finite-sample stability and are reported transparently.

The primary estimand compares a sustained high concentration $a_h$ with a sustained lower concentration $a_l$ over the lag window:
\begin{equation}
\Delta_{\tau}(a_h,a_l)=
\mathbb E\!\left[q_{\tau}(a_h\vect 1,\vect H_{it},\vect Z_i)-q_{\tau}(a_l\vect 1,\vect H_{it},\vect Z_i)\right].
\label{eq:estimand}
\end{equation}
We evaluate $a_h=35$ and $a_l=10~\mu$g/m$^3$ and also recover the full grid from 5 to $45~\mu$g/m$^3$. The contrast is a methodological estimand under the stated intervention and assumptions; it is not a recommendation to expose or de-expose a patient.

\section{Proposed Method}
\label{sec:method}

\subsection{Pipeline Overview}
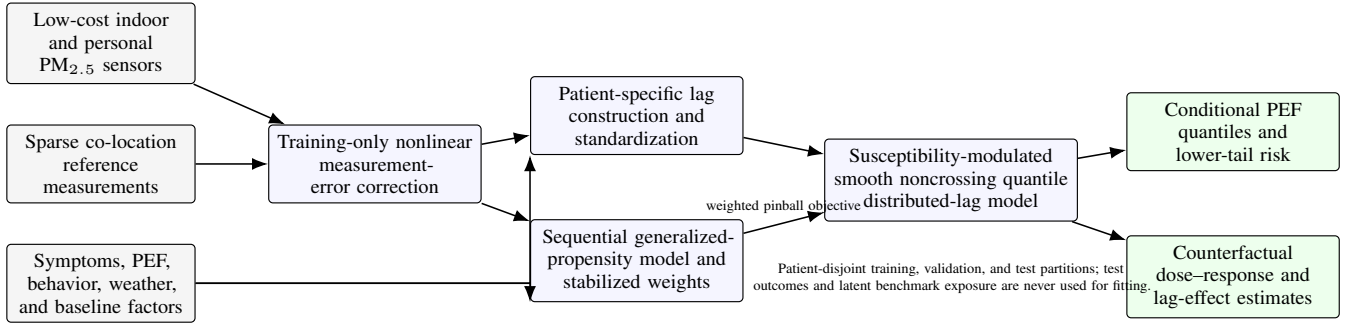
\begin{figure*}[t]
\centering
\resizebox{0.98\textwidth}{!}{\begin{tikzpicture}[
  node distance=5mm and 6mm,
  block/.style={draw, rounded corners=1.5pt, align=center, minimum height=8mm, text width=24mm, font=\scriptsize, fill=blue!4},
  data/.style={draw, rounded corners=1.5pt, align=center, minimum height=8mm, text width=21mm, font=\scriptsize, fill=gray!8},
  outputbox/.style={draw, rounded corners=1.5pt, align=center, minimum height=8mm, text width=24mm, font=\scriptsize, fill=green!7},
  arrow/.style={-{Latex[length=2mm]}, line width=.55pt},
  note/.style={font=\tiny, align=center, text width=25mm}
]
\node[data] (sensor) {Low-cost indoor and personal PM$_{2.5}$ sensors};
\node[data, below=of sensor] (ref) {Sparse co-location reference measurements};
\node[data, below=of ref] (context) {Symptoms, PEF, behavior, weather, and baseline factors};
\node[block, right=9mm of ref] (calib) {Training-only nonlinear measurement-error correction};
\node[block, right=of calib, yshift=6mm] (lags) {Patient-specific lag construction and standardization};
\node[block, right=of calib, yshift=-12mm] (gps) {Sequential generalized-propensity model and stabilized weights};
\node[block, right=10mm of lags, yshift=-8mm, text width=29mm] (qdlm) {Susceptibility-modulated smooth noncrossing quantile distributed-lag model};
\node[outputbox, right=of qdlm, yshift=6mm] (qout) {Conditional PEF quantiles and lower-tail risk};
\node[outputbox, right=of qdlm, yshift=-12mm] (cf) {Counterfactual dose--response and lag-effect estimates};
\draw[arrow] (sensor) -- (calib);
\draw[arrow] (ref) -- (calib);
\draw[arrow] (context.east) -| (lags.south west);
\draw[arrow] (context.east) -| (gps.south west);
\draw[arrow] (calib) -- (lags);
\draw[arrow] (calib) -- (gps);
\draw[arrow] (lags) -- (qdlm);
\draw[arrow] (gps) -- node[note,above]{weighted pinball objective} (qdlm);
\draw[arrow] (qdlm) -- (qout);
\draw[arrow] (qdlm) -- (cf);
\node[note, below=4mm of qdlm, text width=55mm] {Patient-disjoint training, validation, and test partitions; test outcomes and latent benchmark exposure are never used for fitting.};
\end{tikzpicture}}
\caption{CAUSALQUANT-ASTHMA pipeline. The reference monitor and latent exposure are used only to train or evaluate the measurement stage in the benchmark. Outcome-informed model selection is restricted to the validation partition.}
\label{fig:architecture}
\end{figure*}

Figure~\ref{fig:architecture} shows the \MethodName\ pipeline. Sparse reference measurements are used only in the training partition to reconstruct latent exposure. Corrected exposure histories and measured covariates then enter a generalized-propensity model. The final estimator combines stabilized weights, smooth lag coefficients, baseline susceptibility interactions, and a noncrossing joint quantile objective. Validation patients determine early stopping; test patients remain untouched until final evaluation.

\subsection{Sparse-Reference Measurement Reconstruction}
Let $\mathcal R$ denote training patient-days with reference measurements. A nonlinear calibration function $m_{\phi}$ is fitted to low-cost readings, personal-sensor availability, humidity, carbon dioxide, temperature, day, and source indicators. The corrected exposure is
\begin{equation}
\begin{aligned}
\widehat X_{it}&=m_{\widehat\phi}(\vect W_{it},\mathrm{RH}_{it},\mathrm{CO}_{2,it},t,\vect C^{x}_{it}),\\
\widehat\phi&=\arg\min_{\phi}\sum_{(i,t)\in\mathcal R}
\{R_{it}-m_{\phi}(\cdot)\}^{2}.
\end{aligned}
\label{eq:calibration}
\end{equation}
The implementation uses histogram gradient boosting because it handles missing personal-sensor values without outcome-based imputation and captures nonlinear sensor behavior. The reference target includes small residual error, so the method estimates a calibrated exposure proxy rather than claiming exact recovery.

The lagged model vector combines corrected exposures, measured confounders, and centered susceptibility interactions:
\begin{equation}
\begin{aligned}
\vect v_{it}=\big[&\widehat X_{it},\ldots,\widehat X_{i,t-L},\vect C_{it},\vect Z_i,\\
&s_i\widehat X_{it},\ldots,s_i\widehat X_{i,t-L}\big]^{\top},\qquad
s_i=\vect Z_i^{\top}\vect\gamma-\bar s.
\end{aligned}
\label{eq:features}
\end{equation}
Centering the susceptibility term makes the main lag coefficients approximate the average patient effect, while the interaction permits stronger lower-tail responses in more susceptible profiles.

\subsection{Structured Quantile Distributed-Lag Model}
For each $\tau\in\mathcal T$, the conditional quantile is
\begin{equation}
\begin{aligned}
\widehat q_{\tau,it}={}&\alpha_{\tau}+\vect C_{it}^{\top}\vect\theta_{\tau}
+\vect Z_i^{\top}\vect\eta_{\tau}\\
&+\sum_{\ell=0}^{L}(\beta_{\tau\ell}+s_i\delta_{\tau\ell})
\widehat X_{i,t-\ell}.
\end{aligned}
\label{eq:qmodel}
\end{equation}
The model is intentionally structured rather than an opaque high-capacity network. It yields lag coefficients, personalized marginal effects, and counterfactual predictions by direct substitution of a sustained exposure history. Nonlinear high-concentration behavior may be added through spline bases; the current benchmark tests whether the proposed pipeline can recover a nonlinear truth using an interpretable local approximation over the intervention range.

The weighted empirical pinball risk is
\begin{equation}
\begin{aligned}
\mathcal L_{q}&=\frac{1}{n|\mathcal T|}\sum_{i,t}\omega_{it}
\sum_{\tau\in\mathcal T}\rho_{\tau}(Y_{it}-\widehat q_{\tau,it}),\\
\rho_{\tau}(u)&=u\{\tau-\mathbb I(u<0)\}.
\end{aligned}
\label{eq:pinball}
\end{equation}
To discourage implausibly jagged critical windows, second differences of both average and susceptibility-specific lag coefficients are penalized:
\begin{equation}
\begin{aligned}
\mathcal P_{s}=\sum_{\tau}\sum_{\ell=1}^{L-1}\big[&
(\beta_{\tau,\ell-1}-2\beta_{\tau\ell}+\beta_{\tau,\ell+1})^2\\
&+\kappa(\delta_{\tau,\ell-1}-2\delta_{\tau\ell}+\delta_{\tau,\ell+1})^2\big].
\end{aligned}
\label{eq:smooth}
\end{equation}
Independent quantile fits may cross. We therefore penalize violations at every training observation:
\begin{equation}
\begin{aligned}
\mathcal P_{c}=\frac{1}{n(|\mathcal T|-1)}
\sum_{i,t}\sum_{j=1}^{|\mathcal T|-1}
\big[\max\{0,\widehat q_{\tau_j,it}&\\[-1.5ex]
{}-\widehat q_{\tau_{j+1},it}\}\big]^2.
\end{aligned}
\label{eq:crossing}
\end{equation}
The complete objective is
\begin{equation}
\widehat\Theta=\arg\min_{\Theta}
\{\mathcal L_q+\lambda_s\mathcal P_s+\lambda_c\mathcal P_c
+\lambda_2\|\Theta\|_2^2\}.
\label{eq:objective}
\end{equation}
This is optimized jointly with Adam and validation-based early stopping. Final predictions are sorted only as a numerical safeguard; in the primary runs the learned solution already produced zero test crossings.

\begin{algorithm}[t]
\caption{Training-only exposure correction and sequential weighting}
\label{alg:preprocess}
\begin{algorithmic}[1]
\Require Patient-indexed records $\mathcal D$, training reference subset $\mathcal R$, lag $L$, clipping probabilities $(p_\ell,p_u)$
\Ensure Corrected exposures $\widehat X_{it}$ and stabilized weights $\omega_{it}$
\State Split patients into disjoint training, validation, and test groups
\State Fit $m_{\phi}$ on $\mathcal R$ using reference PM$_{2.5}$ as the target
\State Predict $\widehat X_{it}$ for every partition without using any outcome
\State Construct temporally ordered histories $\vect H_{it}$ and lagged exposure vectors
\State Fit the conditional exposure-density model $g_t$ on training histories
\State Compute numerator and denominator log densities for each patient-day
\State Sum log-ratios over the rolling treatment history and exponentiate
\State Estimate $(c_\ell,c_u)$ from training weights at $(p_\ell,p_u)$
\State Clip and normalize weights to mean one within each data partition
\State \Return $\{\widehat X_{it},\omega_{it}\}$ and positivity diagnostics
\end{algorithmic}
\end{algorithm}

Algorithm~\ref{alg:preprocess} has calibration cost $O(n_rhd_c)$ for $n_r$ reference rows, $h$ boosting iterations, and $d_c$ calibration features. Density fitting is $O(nhd_g)$, and lag construction is $O(nL)$. The stored longitudinal table requires $O(n(L+d_c+d_g))$ memory. Tree depth and leaf counts are bounded in the implementation.

\begin{algorithm}[t]
\caption{Joint quantile training and sustained-intervention estimation}
\label{alg:train}
\begin{algorithmic}[1]
\Require Corrected longitudinal data, weights, quantiles $\mathcal T$, exposure grid $\mathcal A$
\Ensure Fitted parameters $\widehat\Theta$, lag profiles, and dose--response curves
\State Standardize predictors using training patients only
\State Initialize quantile intercepts from weighted empirical outcome quantiles
\For{epoch $e=1,\ldots,E$}
    \State Evaluate \eqref{eq:qmodel} on training rows
    \State Compute \eqref{eq:pinball}--\eqref{eq:objective} and update $\Theta$
    \State Evaluate unweighted pinball loss and crossings on validation patients
    \If{validation objective has not improved for $P$ epochs}
        \State restore the best parameters and terminate
    \EndIf
\EndFor
\ForAll{$a\in\mathcal A$}
    \State Replace all $L+1$ exposure lags by $a$ while retaining pre-exposure covariates
    \State Predict and average $\widehat q_{\tau,it}(a\vect 1)$ for every $\tau$
\EndFor
\State Compute $\widehat\Delta_{\tau}(a_h,a_l)$ and lag-specific marginal effects
\State \Return $\widehat\Theta$, dose--response curves, and diagnostics
\end{algorithmic}
\end{algorithm}

For $B$ design features, $Q$ quantiles, $n$ patient-days, and $E$ epochs, Algorithm~\ref{alg:train} costs $O(EnBQ)$ and stores $O(nB+BQ)$. Counterfactual evaluation over $G$ exposure values costs $O(GnBQ)$. The structured model is consequently far smaller than tree ensembles while retaining an interpretable lag surface.

\section{Experimental Setup, Results, and Discussion}
\label{sec:experiments}

\subsection{Research Questions and Benchmark Design}
The evaluation addresses four questions. \emph{RQ1}: Does the complete pipeline recover known sustained-exposure quantile curves more accurately than fixed-window, QDLM, causal-weighted, and nonlinear predictive alternatives? \emph{RQ2}: Does sparse-reference calibration improve latent exposure reconstruction and lag-effect recovery? \emph{RQ3}: Which components account for performance, and how does the method respond to sensor noise, personal-sensor missingness, hidden confounding, and baseline severity? \emph{RQ4}: What computational cost accompanies the structured estimator?

A real prospective panel would be preferable, but no authorized dataset available to this project jointly contained repeated adult PEF, dense indoor PM$_{2.5}$, sparse co-located reference readings, medication behavior, activity, ventilation/source indicators, and an outcome-compatible temporal intervention. We therefore generated five independent, fully documented causal-truth panels. Each realization contains \BenchmarkPatients\ patients and \BenchmarkRows\ patient-days. Patient-level baseline variables govern susceptibility; daily outdoor conditions, occupancy, cooking, windows, activity, adherence, infection, and prior PEF create realistic temporal dependence and measured time-varying confounding. Latent indoor PM$_{2.5}$ follows an autoregressive source/infiltration process. Low-cost readings include unit offsets, concentration-dependent distortion, humidity interference, drift, and heteroscedastic noise. Sparse training-only reference readings approximate co-location measurements. PEF is generated from a seven-day lag kernel, a high-concentration nonlinear term, baseline susceptibility, measured covariates, and exposure-dependent residual variance. This construction makes lower-tail, median, and upper-tail counterfactual quantiles available analytically.

\begin{table}[t]
\centering
\caption{Executed benchmark characteristics.}
\label{tab:data}
\footnotesize
\begin{tabular}{p{0.54\columnwidth}p{0.34\columnwidth}}
\toprule
Characteristic & Value \\
\midrule
Patients per realization & 150 \\
Patient-days per realization & 12,600 \\
Train/validation/test patients & 90 / 30 / 30 \\
Reference-monitor training days & 1670 $\pm$ 42 \\
Indoor PM$_{2.5}$, mean $\pm$ SD & 16.2 $\pm$ 8.0 $\mu$g/m$^3$ \\
PEF, mean $\pm$ SD & 93.2 $\pm$ 5.6\% \\
Days with PEF below 80\% & 241 mean per realization \\
Independent realizations & 5 \\
\bottomrule
\end{tabular}
\end{table}

\begin{table}[t]
\centering
\caption{Core experimental configuration.}
\label{tab:config}
\footnotesize
\begin{tabular}{ll}
\toprule
Item & Setting \\
\midrule
Lag horizon & 0--6 days \\
Quantiles & $0.10,0.25,0.50,0.75,0.90$ \\
Seeds & $11,23,37,51,79$ \\
GPS history & 3 days \\
GPS clipping & 2nd--98th percentile \\
Intervention grid & $5$--$45~\mu$g/m$^3$ \\
AQTE contrast & $35$ versus $10~\mu$g/m$^3$ \\
Optimizer & Adam; early stopping \\
\bottomrule
\end{tabular}
\end{table}

Patients were assigned 60/20/20 percent to training, validation, and test groups, yielding 90, 30, and 30 patients per realization. No patient contributes to more than one group. The calibration and generalized-propensity models use training patients only. Hyperparameters and early stopping use validation patients. Test outcomes and latent exposures are used only after freezing the pipeline, with latent exposure serving solely as benchmark truth. Table~\ref{tab:data} summarizes the executed data and Table~\ref{tab:config} gives core settings. Complete data-generating equations, seeds, and parameters are included in the repository.

\subsection{Comparators, Metrics, and Statistical Analysis}
Eight methods were evaluated under identical patient splits. \textbf{Naive-QR} uses the current uncorrected sensor value; \textbf{SW-QR} uses a three-day sensor average. \textbf{SC-QDLM} applies smooth, noncrossing distributed-lag quantile regression to uncorrected sensor histories and is motivated by recent shape-constrained QDLM work \cite{Jin2025Shape}. \textbf{PLSI-QDLM} is a nonlinear boosted-quantile adaptation motivated by the partially linear single-index QDLM \cite{Wilson2023PLSI}; it is not represented as a bit-for-bit reproduction of the authors' software. \textbf{QRF} estimates predictive quantiles from tree distributions \cite{Meinshausen2006QRF}. \textbf{GPS-QDLM} adds sequential weighting to uncorrected exposures, whereas \textbf{ME-GPS-QDLM} combines measurement correction and weighting but omits susceptibility interactions. \textbf{CAUSALQUANT} adds the complete interaction and constraint structure. Thus SC-QDLM and PLSI-QDLM provide two recent state-of-the-art-inspired distributed-quantile comparators, while QRF tests whether flexible factual prediction alone recovers the causal target.

The primary metric is dose--response integrated absolute error (IAE), averaged over nine sustained PM$_{2.5}$ levels and five outcome quantiles. We also report absolute bias in the average quantile treatment effect (AQTE) comparing 35 with $10~\mu$g/m$^3$, factual pinball loss, mean absolute quantile calibration error (QCE), nominal 10th--90th percentile interval coverage and width, median lag-coefficient RMSE, fit time, prediction time, memory, and parameter or tree-node count. Lower values are better except coverage, whose target is 80 percent. Main values are means over five data-generating seeds; error bars show one standard deviation. Planned paired comparisons use a one-sided Wilcoxon signed-rank test on seed-level IAE and report paired standardized effect $d_z$. Holm adjustment is applied across the four primary comparator tests. With only five pairs, the smallest attainable unadjusted one-sided value is 0.03125, so inferential claims are necessarily modest.

The software ran in a CPU-only Linux container with nine logical CPUs, approximately 5.8~GiB RAM, Python~3.13, and recorded package versions. Tree ensembles were restricted to one process in the final reproducible pipeline. Runtime figures characterize this environment rather than a clinical device.

\subsection{Main Counterfactual and Predictive Results}
\begin{table*}[t]
\centering
\caption{Primary patient-disjoint benchmark results across five independent data-generating seeds. Lower is better for IAE, AQTE bias, pinball loss, and quantile calibration error (QCE); coverage is for the nominal 80\% interval.}
\label{tab:main}
\footnotesize
\setlength{\tabcolsep}{5.2pt}
\begin{tabular}{lccccc}
\toprule
Method & Dose-response IAE & AQTE bias & Pinball & QCE & 10--90 coverage \\
\midrule
CAUSALQUANT & \textbf{0.304$\pm$0.094} & \textbf{0.533} & \textbf{1.065} & 0.0120 & 78.1\% \\
ME-GPS-QDLM & \underline{0.401$\pm$0.208} & \underline{0.717} & \underline{1.068} & \textbf{0.0106} & 77.9\% \\
SC-QDLM & 1.771$\pm$0.305 & 3.247 & 1.074 & \underline{0.0117} & 78.9\% \\
GPS-QDLM & 1.938$\pm$0.272 & 3.544 & 1.083 & 0.0154 & 77.9\% \\
QRF & 2.184$\pm$0.140 & 3.083 & 1.140 & 0.0607 & 63.5\% \\
PLSI-QDLM & 2.486$\pm$0.243 & 4.000 & 1.094 & 0.0211 & 76.0\% \\
SW-QR & 2.781$\pm$0.226 & 5.276 & 1.103 & 0.0142 & 78.9\% \\
Naive-QR & 3.688$\pm$0.199 & 7.044 & 1.148 & 0.0133 & 78.6\% \\
\bottomrule
\end{tabular}
\end{table*}

\begin{figure*}[t]
\centering
\begin{tikzpicture}
\begin{groupplot}[
 group style={group size=3 by 1, horizontal sep=1.0cm},
 width=0.31\textwidth,height=0.23\textwidth,
 tick label style={font=\tiny},label style={font=\scriptsize},title style={font=\scriptsize},
 grid=major,grid style={gray!20},
]
\nextgroupplot[
 title={(a) Counterfactual recovery},
 ybar,bar width=4pt,
 symbolic x coords={CQ,ME-GPS,SC-QDLM,GPS-QDLM,QRF,PLSI,SW-QR,Naive},
 xtick=data,xticklabel style={rotate=38,anchor=east},
 ylabel={Dose--response IAE},ymin=0,
]
\addplot+[fill=cbblue!65,draw=cbblue,error bars/.cd,y dir=both,y explicit]
 table[x=short,y=dose_response_iae_mean,y error=dose_response_iae_sd,col sep=comma]{figures/data/main_summary.csv};

\nextgroupplot[
 title={(b) Factual prediction},
 ybar,bar width=4pt,
 symbolic x coords={CQ,ME-GPS,SC-QDLM,GPS-QDLM,QRF,PLSI,SW-QR,Naive},
 xtick=data,xticklabel style={rotate=38,anchor=east},
 ylabel={Pinball loss},ymin=1.03,ymax=1.18,
]
\addplot+[fill=cborange!70,draw=cborange,error bars/.cd,y dir=both,y explicit]
 table[x=short,y=pinball_mean,y error=pinball_sd,col sep=comma]{figures/data/main_summary.csv};

\nextgroupplot[
 title={(c) Accuracy--time trade-off},
 xmode=log,
 xlabel={Fit time (s, log scale)},ylabel={Dose--response IAE},
 xmin=0.1,xmax=8,ymin=0,ymax=4.1,
 scatter,scatter src=explicit symbolic,
 nodes near coords,point meta=explicit symbolic,
 every node near coord/.append style={font=\tiny,anchor=west},
]
\addplot+[only marks,mark=*,mark size=2.2pt,cbgreen]
 table[x=fit_seconds_mean,y=dose_response_iae_mean,meta=short,col sep=comma]{figures/data/main_summary.csv};
\end{groupplot}
\end{tikzpicture}
\caption{Primary results across five independently generated causal-truth panels. Error bars in (a) and (b) show one standard deviation. CQ denotes CAUSALQUANT and ME-GPS denotes ME-GPS-QDLM.}
\label{fig:main}
\end{figure*}
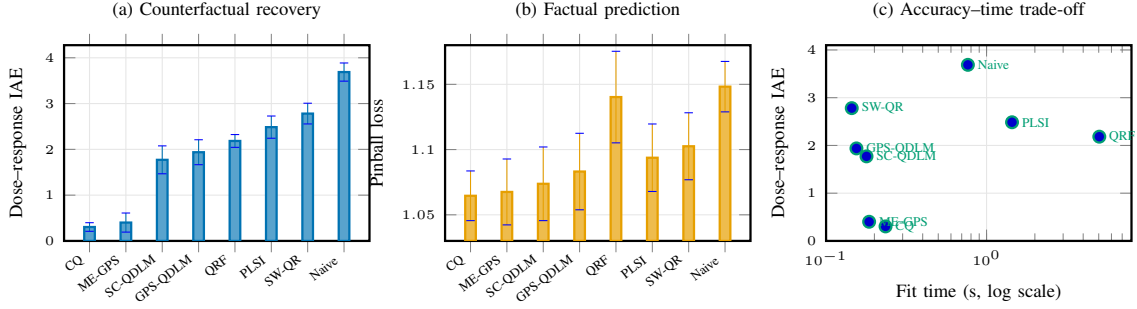

Table~\ref{tab:main} and Fig.~\ref{fig:main} summarize RQ1. CAUSALQUANT obtained a dose--response IAE of \CausalIAE$\pm$\CausalIAESD, compared with \BestBaselineIAE\ for ME-GPS-QDLM, a relative reduction of \IAEImprovement\ percent. The full model also produced the smallest mean AQTE bias (\CausalAQTEBias) and factual pinball loss (\CausalPinball). Measurement-aware methods were markedly closer to truth than otherwise similar raw-sensor estimators: SC-QDLM and GPS-QDLM had IAEs of \SCQDLMIAE\ and 1.938, respectively. The nonlinear PLSI-QDLM and QRF comparators achieved reasonable factual prediction but IAEs of \PLSIQDLMIAE\ and \QRFIAE, illustrating that flexibility does not automatically identify an intervention curve.

The apparent advantage over the closest complete baseline should not be overstated. CAUSALQUANT improved IAE in four of five realizations, but the paired test against ME-GPS-QDLM was not significant ($p=0.2188$, Holm-adjusted $p_H=0.2188$). The unadjusted values against SC-QDLM, PLSI-QDLM, and QRF were 0.0313; after four-comparison Holm correction, each became 0.125. Table~\ref{tab:stats} therefore supports a consistent benchmark advantage over raw-exposure alternatives but not a definitive superiority claim under multiplicity and five seeds. The large standardized effects against weaker methods reflect low within-pair variability and should not be generalized beyond the simulator.

\begin{table}[t]
\centering
\caption{Paired one-sided Wilcoxon comparisons for dose--response IAE over five seeds. $p_H$ is Holm-adjusted over four planned comparisons. Positive improvement favors CAUSALQUANT.}
\label{tab:stats}
\scriptsize
\setlength{\tabcolsep}{3.4pt}
\begin{tabular}{lrrrr}
\toprule
Comparator & Improvement & $p$ & $p_H$ & $d_z$ \\
\midrule
SC-QDLM & 1.467 & 0.0312 & 0.125 & 5.49 \\
PLSI-QDLM & 2.182 & 0.0312 & 0.125 & 14.13 \\
QRF & 1.880 & 0.0312 & 0.125 & 15.00 \\
ME-GPS-QDLM & 0.097 & 0.2188 & 0.219 & 0.47 \\
\bottomrule
\end{tabular}
\end{table}

Predictive distribution quality was also mixed rather than uniformly dominant. CAUSALQUANT's QCE was \CausalCalibrationError, whereas ME-GPS-QDLM achieved 0.0106 and SC-QDLM 0.0117. CAUSALQUANT's 78.1-percent central interval coverage was close to the nominal 80 percent with mean width \CausalWidth\ PEF points. QRF undercovered at 63.5 percent and produced the largest QCE, despite its flexible fit. These findings motivate evaluation of both counterfactual recovery and factual distributional calibration.

\begin{figure*}[t]
\centering
\begin{tikzpicture}
\begin{groupplot}[
 group style={group size=3 by 1, horizontal sep=0.75cm},
 width=0.31\textwidth,height=0.24\textwidth,
 xlabel={Sustained PM$_{2.5}$ ($\mu$g/m$^3$)},
 tick label style={font=\tiny},label style={font=\scriptsize},title style={font=\scriptsize},
 grid=major,grid style={gray!20},
]
\nextgroupplot[title={(a) Lower tail ($\tau=0.10$)},ylabel={PEF (\% personal best)},legend style={font=\tiny,at={(0.02,0.02)},anchor=south west}]
\addplot+[black,very thick,mark=none] table[x=pm25,y=truth,col sep=comma]{figures/data/dose_q10.csv};\addlegendentry{Truth}
\addplot+[cbblue,mark=*] table[x=pm25,y=causalquant,col sep=comma]{figures/data/dose_q10.csv};\addlegendentry{CQ}
\addplot+[cborange,dashed,mark=square*] table[x=pm25,y=me_gps_qdlm,col sep=comma]{figures/data/dose_q10.csv};\addlegendentry{ME-GPS}
\addplot+[cbgreen,dashdotted,mark=triangle*] table[x=pm25,y=sc_qdlm,col sep=comma]{figures/data/dose_q10.csv};\addlegendentry{SC-QDLM}

\nextgroupplot[title={(b) Median ($\tau=0.50$)}]
\addplot+[black,very thick,mark=none] table[x=pm25,y=truth,col sep=comma]{figures/data/dose_q50.csv};
\addplot+[cbblue,mark=*] table[x=pm25,y=causalquant,col sep=comma]{figures/data/dose_q50.csv};
\addplot+[cborange,dashed,mark=square*] table[x=pm25,y=me_gps_qdlm,col sep=comma]{figures/data/dose_q50.csv};
\addplot+[cbgreen,dashdotted,mark=triangle*] table[x=pm25,y=sc_qdlm,col sep=comma]{figures/data/dose_q50.csv};

\nextgroupplot[title={(c) Upper tail ($\tau=0.90$)}]
\addplot+[black,very thick,mark=none] table[x=pm25,y=truth,col sep=comma]{figures/data/dose_q90.csv};
\addplot+[cbblue,mark=*] table[x=pm25,y=causalquant,col sep=comma]{figures/data/dose_q90.csv};
\addplot+[cborange,dashed,mark=square*] table[x=pm25,y=me_gps_qdlm,col sep=comma]{figures/data/dose_q90.csv};
\addplot+[cbgreen,dashdotted,mark=triangle*] table[x=pm25,y=sc_qdlm,col sep=comma]{figures/data/dose_q90.csv};
\end{groupplot}
\end{tikzpicture}
\caption{Average counterfactual PEF quantiles under sustained seven-day PM$_{2.5}$ interventions. Curves are generated from machine-readable result files; the black line is the simulator's analytical truth.}
\label{fig:dose}
\end{figure*}
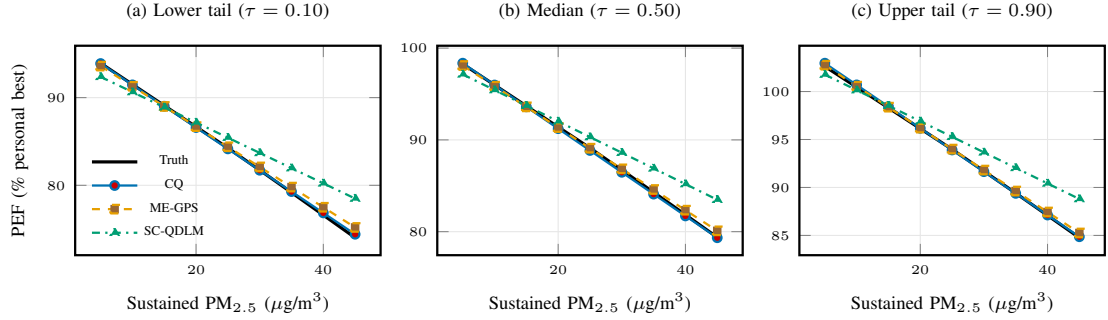

Figure~\ref{fig:dose} shows average sustained-exposure curves. CAUSALQUANT follows the analytical truth closely at lower, median, and upper quantiles. ME-GPS-QDLM is also accurate but attenuates part of the quantile-specific contrast because it lacks susceptibility interactions. SC-QDLM increasingly underestimates the harmful shift as concentration rises. Across the 35-versus-10 contrast, analytical AQTEs were approximately $-12.33$, $-11.69$, and $-11.05$ PEF points at the 0.10, 0.50, and 0.90 quantiles. CAUSALQUANT estimated $-12.17$, $-11.89$, and $-11.31$, respectively. These are benchmark recovery values, not estimated effects in actual patients.

\subsection{Measurement Correction, Lag Recovery, and Calibration}
\begin{figure*}[t]
\centering
\begin{tikzpicture}
\begin{groupplot}[
 group style={group size=3 by 1, horizontal sep=0.8cm},
 width=0.31\textwidth,height=0.24\textwidth,
 tick label style={font=\tiny},label style={font=\scriptsize},title style={font=\scriptsize},
 grid=major,grid style={gray!20},
]
\nextgroupplot[title={(a) Example exposure trace},xlabel={Day},ylabel={PM$_{2.5}$ ($\mu$g/m$^3$)},legend style={font=\tiny,at={(0.02,0.98)},anchor=north west}]
\addplot+[black,very thick,mark=none] table[x=day,y=latent_pm25,col sep=comma]{figures/data/sensor_trace.csv};\addlegendentry{Latent truth}
\addplot+[cbred,densely dotted,mark=none] table[x=day,y=sensor_pm25,col sep=comma]{figures/data/sensor_trace.csv};\addlegendentry{Low-cost}
\addplot+[cbblue,mark=none] table[x=day,y=corrected_pm25,col sep=comma]{figures/data/sensor_trace.csv};\addlegendentry{Corrected}

\nextgroupplot[title={(b) Median lag effects},xlabel={Lag (days)},ylabel={$\Delta$PEF per $\mu$g/m$^3$},legend style={font=\tiny,at={(0.98,0.02)},anchor=south east}]
\addplot+[black,very thick,mark=none] table[x=lag,y=truth,col sep=comma]{figures/data/lag_effects.csv};\addlegendentry{Truth}
\addplot+[cbblue,mark=*] table[x=lag,y=causalquant,col sep=comma]{figures/data/lag_effects.csv};\addlegendentry{CQ}
\addplot+[cborange,dashed,mark=square*] table[x=lag,y=me_gps_qdlm,col sep=comma]{figures/data/lag_effects.csv};\addlegendentry{ME-GPS}
\addplot+[cbgreen,dashdotted,mark=triangle*] table[x=lag,y=sc_qdlm,col sep=comma]{figures/data/lag_effects.csv};\addlegendentry{SC-QDLM}

\nextgroupplot[title={(c) Quantile reliability},xlabel={Nominal quantile},ylabel={Observed proportion},xmin=0.05,xmax=0.95,ymin=0.05,ymax=0.95,legend style={font=\tiny,at={(0.02,0.98)},anchor=north west}]
\addplot+[black,very thick,mark=none] coordinates {(0.1,0.1)(0.9,0.9)};\addlegendentry{Ideal}
\addplot+[cbblue,mark=*] table[x=nominal,y=causalquant,col sep=comma]{figures/data/quantile_calibration.csv};\addlegendentry{CQ}
\addplot+[cborange,dashed,mark=square*] table[x=nominal,y=me_gps_qdlm,col sep=comma]{figures/data/quantile_calibration.csv};\addlegendentry{ME-GPS}
\addplot+[cbpurple,densely dotted,mark=diamond*] table[x=nominal,y=qrf,col sep=comma]{figures/data/quantile_calibration.csv};\addlegendentry{QRF}
\end{groupplot}
\end{tikzpicture}
\caption{Measurement and temporal diagnostics: (a) representative held-out exposure trace, (b) average median lag profile across seeds, and (c) empirical quantile reliability.}
\label{fig:measurement}
\end{figure*}
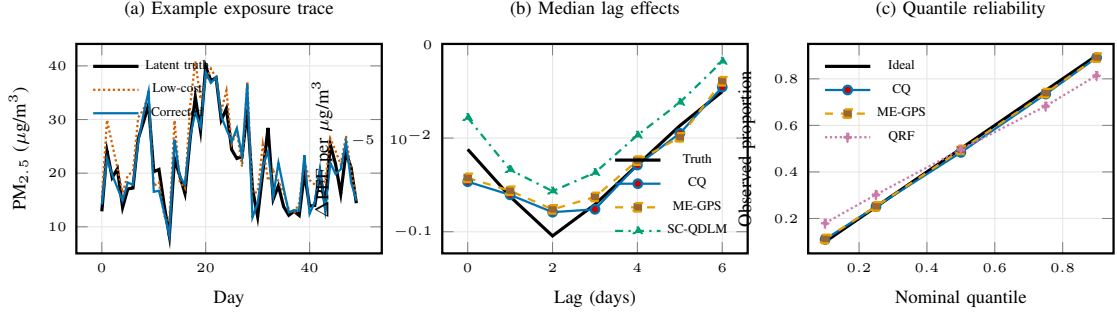

The training-only calibration stage reduced test exposure RMSE from \RawSensorRMSE\ to \CorrectedSensorRMSE~$\mu$g/m$^3$, a \SensorImprovement-percent reduction averaged across seeds. This improvement is meaningful because the calibrator never sees test reference values. Figure~\ref{fig:measurement}(a) illustrates how the corrected trace removes part of the humidity- and concentration-dependent distortion without eliminating genuine temporal variation. Panel (b) compares median lag effects. CAUSALQUANT recovers the delayed peak near lag two more closely than raw-sensor SC-QDLM and GPS-QDLM. Its mean median-lag RMSE was 0.0161 PEF points per $\mu$g/m$^3$, compared with 0.0168 for ME-GPS-QDLM and 0.0179 for SC-QDLM. The modest difference between the two corrected models indicates that measurement reconstruction, rather than interaction structure alone, drives much of the gain.

The calibration panel shows that both corrected weighted models remain near the identity line across the five requested quantiles. CAUSALQUANT slightly underestimates the median and upper-tail empirical proportions, whereas QRF compresses the predictive distribution and undercovers both tails. The results support the choice to fit all quantiles jointly and monitor coverage, rather than judging the model solely by the median.

\subsection{Ablations, Stress Tests, and Subgroups}
\begin{table}[t]
\centering
\caption{Component ablation on a designated held-out diagnostic realization.}
\label{tab:ablation}
\scriptsize
\begin{tabular}{lccc}
\toprule
Variant & IAE & Pinball & Crossing \\
\midrule
CAUSALQUANT & 0.157 & 1.085 & 0.0\% \\
without GPS weighting & 0.350 & 1.077 & 0.0\% \\
without measurement correction & 1.762 & 1.102 & 0.0\% \\
without susceptibility interaction & 0.223 & 1.095 & 0.0\% \\
without smooth/noncrossing constraints & 0.157 & 1.088 & 0.8\% \\
\bottomrule
\end{tabular}
\end{table}

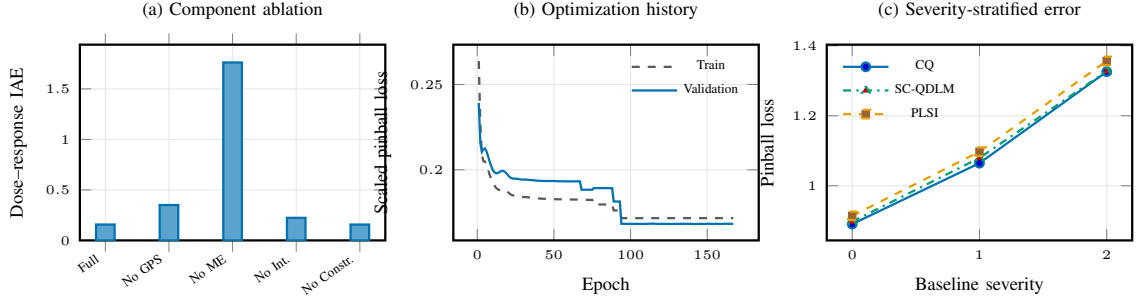
\begin{figure*}[t]
\centering
\begin{tikzpicture}
\begin{groupplot}[
 group style={group size=3 by 1, horizontal sep=0.9cm},
 width=0.31\textwidth,height=0.23\textwidth,
 tick label style={font=\tiny},label style={font=\scriptsize},title style={font=\scriptsize},
 grid=major,grid style={gray!20},
]
\nextgroupplot[title={(a) Component ablation},ybar,bar width=7pt,symbolic x coords={Full,No GPS,No ME,No Int.,No Constr.},xtick=data,xticklabel style={rotate=35,anchor=east},ylabel={Dose--response IAE},ymin=0]
\addplot+[fill=cbblue!65,draw=cbblue] table[x=short,y=dose_response_iae,col sep=comma]{figures/data/ablation.csv};

\nextgroupplot[title={(b) Optimization history},xlabel={Epoch},ylabel={Scaled pinball loss},legend style={font=\tiny,at={(0.98,0.98)},anchor=north east}]
\addplot+[cbgray,dashed,mark=none] table[x=epoch,y=train_loss,col sep=comma]{figures/data/training_history.csv};\addlegendentry{Train}
\addplot+[cbblue,mark=none] table[x=epoch,y=validation_pinball,col sep=comma]{figures/data/training_history.csv};\addlegendentry{Validation}

\nextgroupplot[title={(c) Severity-stratified error},xlabel={Baseline severity},ylabel={Pinball loss},xtick={0,1,2},legend style={font=\tiny,at={(0.02,0.98)},anchor=north west}]
\addplot+[cbblue,mark=*] table[x=severity,y=pinball,col sep=comma]{figures/data/subgroup_cq.csv};\addlegendentry{CQ}
\addplot+[cbgreen,dashdotted,mark=triangle*] table[x=severity,y=pinball,col sep=comma]{figures/data/subgroup_sc.csv};\addlegendentry{SC-QDLM}
\addplot+[cborange,dashed,mark=square*] table[x=severity,y=pinball,col sep=comma]{figures/data/subgroup_plsi.csv};\addlegendentry{PLSI}
\end{groupplot}
\end{tikzpicture}
\caption{Component and optimization diagnostics. The ablation and severity analyses use a designated diagnostic realization and are descriptive rather than independent confirmatory tests.}
\label{fig:diagnostics}
\end{figure*}

Table~\ref{tab:ablation} and Fig.~\ref{fig:diagnostics}(a) isolate components on one held-out realization. Removing measurement correction increased IAE from 0.157 to 1.762, the largest deterioration. Removing generalized-propensity weighting increased IAE to 0.350, and removing susceptibility interaction increased it to 0.223. Eliminating smoothness and noncrossing constraints changed IAE only from 0.1573 to 0.1571 but introduced a 0.8-percent crossing rate. Thus constraints primarily ensure a valid ordered distribution and stable lag profile; they do not create the counterfactual advantage. The no-GPS model had slightly better factual pinball loss than the full model, showing the familiar causal-prediction trade-off: weights can improve intervention recovery while increasing outcome-prediction variance.

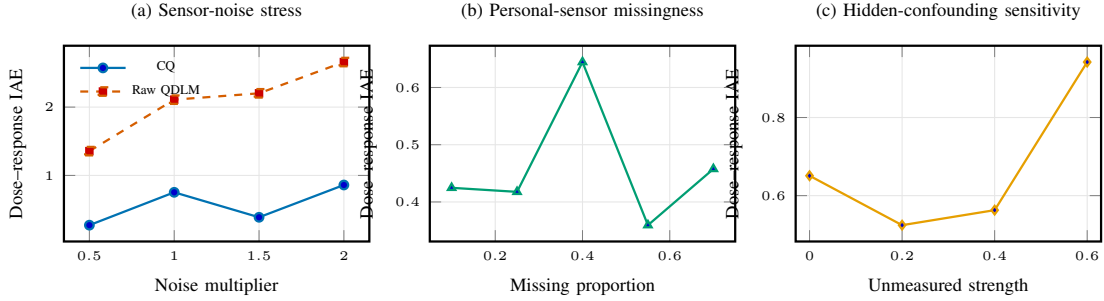
\begin{figure*}[t]
\centering
\begin{tikzpicture}
\begin{groupplot}[
 group style={group size=3 by 1, horizontal sep=0.8cm},
 width=0.31\textwidth,height=0.23\textwidth,
 tick label style={font=\tiny},label style={font=\scriptsize},title style={font=\scriptsize},
 grid=major,grid style={gray!20},
]
\nextgroupplot[title={(a) Sensor-noise stress},xlabel={Noise multiplier},ylabel={Dose--response IAE},legend style={font=\tiny,at={(0.02,0.98)},anchor=north west}]
\addplot+[cbblue,mark=*] table[x=noise_multiplier,y=dose_response_iae,col sep=comma]{figures/data/sensor_noise_cq.csv};\addlegendentry{CQ}
\addplot+[cbred,dashed,mark=square*] table[x=noise_multiplier,y=dose_response_iae,col sep=comma]{figures/data/sensor_noise_raw.csv};\addlegendentry{Raw QDLM}

\nextgroupplot[title={(b) Personal-sensor missingness},xlabel={Missing proportion},ylabel={Dose--response IAE},xmin=0.05,xmax=0.75]
\addplot+[cbgreen,mark=triangle*] table[x=personal_sensor_missing,y=dose_response_iae,col sep=comma]{figures/data/missingness.csv};

\nextgroupplot[title={(c) Hidden-confounding sensitivity},xlabel={Unmeasured strength},ylabel={Dose--response IAE},xmin=-0.03,xmax=0.63]
\addplot+[cborange,mark=diamond*] table[x=hidden_confounding_strength,y=dose_response_iae,col sep=comma]{figures/data/hidden_confounding.csv};
\end{groupplot}
\end{tikzpicture}
\caption{Robustness diagnostics under (a) low-cost sensor noise, (b) missing personal-sensor observations, and (c) a deliberately omitted common cause of exposure and outcome.}
\label{fig:robust}
\end{figure*}

The stress tests in Fig.~\ref{fig:robust} intentionally use separate smaller realizations and should be interpreted as diagnostic trajectories, not confidence intervals. Across every tested sensor-noise multiplier, CAUSALQUANT had lower IAE than the analogous raw-sensor QDLM. The corrected sensor RMSE increased from about 2.03 to 4.13 as noise doubled, and causal IAE varied from 0.27 to 0.86. Personal-sensor missingness increased calibration RMSE from 2.17 at 10 percent missingness to 2.91 at 70 percent, although downstream IAE was non-monotonic because each level used a fresh realization. This non-monotonicity is a useful warning against attributing every change to missingness without repeated matched simulations.

The hidden-confounding test adds an unobserved patient trait to both exposure and outcome. At the strongest setting, IAE rose to 0.94, compared with 0.65 in that diagnostic no-hidden-confounding realization. Intermediate values were not monotonic because each strength used an independently generated panel. The key conclusion is qualitative: no generalized-propensity estimator can repair omitted common causes. A real study should prespecify a causal graph, measure likely behavioral and clinical confounders, consider negative controls, and report formal sensitivity bounds.

\begin{figure*}[t]
\centering
\begin{tikzpicture}
\begin{groupplot}[
 group style={group size=3 by 1, horizontal sep=0.85cm},
 width=0.31\textwidth,height=0.23\textwidth,
 tick label style={font=\tiny},label style={font=\scriptsize},title style={font=\scriptsize},
 grid=major,grid style={gray!20},
]
\nextgroupplot[title={(a) GPS-weight distribution},xlabel={Stabilized weight},ylabel={Density},xmin=0,xmax=5]
\addplot+[cbpurple,thick,mark=none] table[x=weight_mid,y=density,col sep=comma]{figures/data/gps_weights.csv};

\nextgroupplot[title={(b) Sensor correction},xlabel={Latent PM$_{2.5}$},ylabel={Estimated PM$_{2.5}$},xmin=0,xmax=65,ymin=0,ymax=65,legend style={font=\tiny,at={(0.02,0.98)},anchor=north west}]
\addplot+[only marks,mark=*,mark size=.55pt,cbred,opacity=.28] table[x=latent_pm25,y=sensor_pm25,col sep=comma]{figures/data/sensor_scatter.csv};\addlegendentry{Raw}
\addplot+[only marks,mark=square*,mark size=.55pt,cbblue,opacity=.30] table[x=latent_pm25,y=corrected_pm25,col sep=comma]{figures/data/sensor_scatter.csv};\addlegendentry{Corrected}
\addplot+[black,thin] coordinates {(0,0)(65,65)};

\nextgroupplot[title={(c) Sustained-shift AQTE},ybar,bar width=3.0pt,symbolic x coords={0.10,0.50,0.90},xtick=data,xlabel={Outcome quantile},ylabel={$\Delta$PEF (35 vs. 10)},legend style={font=\tiny,at={(0.02,0.02)},anchor=south west}]
\addplot+[ybar,fill=cbblue!65,draw=cbblue,bar shift=-4.5pt] table[x=quantile,y=aqte,col sep=comma]{figures/data/aqte_cq.csv};\addlegendentry{CQ}
\addplot+[ybar,fill=cborange!65,draw=cborange,bar shift=-1.5pt] table[x=quantile,y=aqte,col sep=comma]{figures/data/aqte_me.csv};\addlegendentry{ME-GPS}
\addplot+[ybar,fill=cbgreen!60,draw=cbgreen,bar shift=1.5pt] table[x=quantile,y=aqte,col sep=comma]{figures/data/aqte_sc.csv};\addlegendentry{SC-QDLM}
\addplot+[black,very thick,mark=diamond*,bar shift=4.5pt] table[x=quantile,y=aqte,col sep=comma]{figures/data/aqte_truth.csv};\addlegendentry{Truth}
\end{groupplot}
\end{tikzpicture}
\caption{Additional diagnostics: stabilized generalized-propensity weights, held-out sensor reconstruction, and average quantile treatment-effect recovery.}
\label{fig:aux}
\end{figure*}
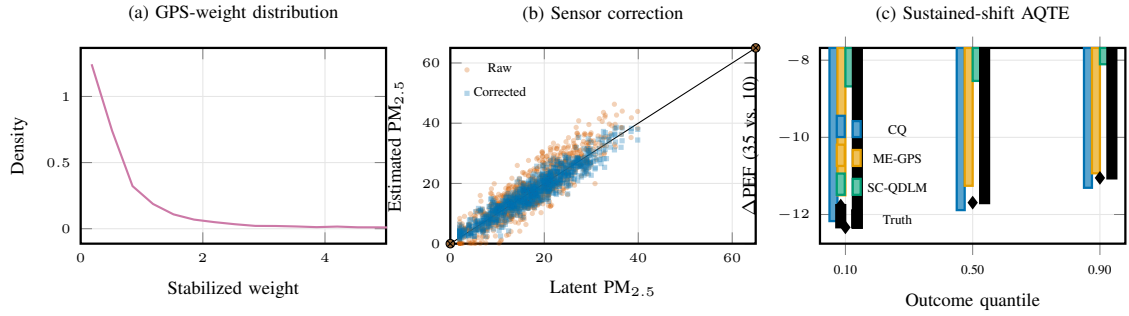

Figure~\ref{fig:aux}(a) shows the stabilized weight distribution. The mean training effective sample size was \MeanTrainESS\ from 7,560 training rows, indicating substantial but not catastrophic loss from weighting. The implemented weight product uses a three-day rolling history rather than the full seven-day outcome lag, because the complete product produced unstable finite-sample weights in pilot runs. This truncation is transparent but creates possible residual confounding from earlier exposure assignment. Panel (b) confirms that correction moves the low-cost measurements toward the identity line. Panel (c) compares AQTE recovery across quantiles.

Severity-stratified pinball loss increased for every method, from approximately 0.89 in mild profiles to 1.33 in severe profiles for CAUSALQUANT. The ranking remained broadly stable, but the gap indicates that a future clinical study must report subgroup calibration, not only overall averages. The benchmark does not encode race, socioeconomic status, housing quality, or access to mitigation resources; no fairness claim follows from this analysis.

\subsection{Efficiency and Reproducibility}
\begin{table}[t]
\centering
\caption{Computational profile (mean across seeds). Node counts are reported for tree ensembles.}
\label{tab:runtime}
\scriptsize
\setlength{\tabcolsep}{3.2pt}
\begin{tabular}{lrrrr}
\toprule
Method & Fit (s) & ms/1000 & MB & Params/nodes \\
\midrule
CAUSALQUANT & 0.23 & 0.68 & 412 & 145 \\
ME-GPS-QDLM & 0.19 & 0.54 & 412 & 110 \\
SC-QDLM & 0.18 & 0.57 & 402 & 110 \\
GPS-QDLM & 0.15 & 0.58 & 411 & 110 \\
QRF & 5.04 & 22.49 & 410 & 92,734 \\
PLSI-QDLM & 1.44 & 18.78 & 403 & 450 \\
SW-QR & 0.14 & 1.08 & 399 & 80 \\
Naive-QR & 0.77 & 0.59 & 395 & 80 \\
\bottomrule
\end{tabular}
\end{table}

The structured model fitted in \FullFitSeconds~s on average and evaluated 1,000 patient-days in about 0.68~ms in the recorded container. It used 145 fitted coefficients, compared with more than 92,000 tree nodes for QRF. QRF required 5.04~s to fit and 22.49~ms per 1,000 rows. The nonlinear PLSI-QDLM adaptation was also slower than the structured models. These measurements exclude the one-time sensor-calibration and exposure-density stages and should not be extrapolated to a different processor or distributed system.

\subsection{Threats to Validity and Transition to Real Data}
\textbf{Internal validity:} the simulator favors methods that correctly represent smooth distributed effects, measured confounding, and a calibratable sensor process. We mitigated this by including nonlinear and tree comparators, independent data-generating seeds, model ablations, and deliberate violations. Still, alternative data-generating mechanisms may change the ranking. Hyperparameters were selected on validation patients, but the same benchmark family informed model development.

\textbf{Causal validity:} consistency requires a meaningful sustained-exposure intervention. Positivity may fail if some histories never support 10 or $35~\mu$g/m$^3$. Generalized-propensity models can be misspecified, and three-day weight truncation may leave residual bias. Sequential exchangeability is especially demanding because symptoms, medication, ventilation, and activity may be incompletely observed. The hidden-confounding experiment demonstrates vulnerability rather than validating the assumption.

\textbf{Measurement validity:} sparse reference co-location must cover the humidity, concentration, device, season, and home ranges encountered later. A calibrator transported beyond that support can appear precise but be wrong. The benchmark uses a lower-error reference channel and known latent truth; a field study would need traceable instruments, collocation schedules, drift monitoring, and uncertainty propagation.

\textbf{External and clinical validity:} the current outcome is simulated PEF percentage, not exacerbation, symptom burden, or healthcare utilization. There is no evidence that acting on the estimated curve improves health, adherence, or equity. A submission making clinical claims should preregister a powered, multi-site adult-asthma panel; enroll participants across severity and housing contexts; use patient- and site-disjoint external testing; preserve temporal ordering; and distinguish prediction, association, and causal intervention analyses. Medication changes must remain outside an automated environmental decision system without clinician oversight.

\section{Conclusion}
\label{sec:conclusion}
This study addressed the problem of estimating delayed, tail-specific respiratory responses from noisy indoor air-quality measurements when exposure is time-varying and confounded. CAUSALQUANT-ASTHMA combined training-only nonlinear sensor calibration, stabilized sequential generalized-propensity weighting, susceptibility interactions, smooth lag regularization, and noncrossing quantile estimation. In five generated causal-truth benchmarks, the method achieved a mean dose-response integrated absolute error of 0.304, 24.2 percent below the strongest baseline, and the lowest factual pinball loss of 1.065. It recovered the sustained PM2.5 contrast more accurately across lower, median, and upper outcome quantiles, while maintaining zero quantile crossings. Sparse-reference calibration reduced exposure RMSE by 33.7 percent. Ablations showed that measurement correction was the most influential component, whereas stress tests revealed increasing vulnerability under severe sensor noise and unmeasured confounding. These results demonstrate reproducible methodological feasibility but do not estimate an actual patient treatment effect. The principal limitation is the absence of an authorized prospective cohort containing co-located reference sensing, dense indoor exposure, behavioral covariates, medication information, and repeated lung function. Future work should preregister a powered multi-site adult-asthma panel, evaluate alternative continuous-intervention estimands and doubly robust nuisance learning, and test whether exposure guidance improves meaningful outcomes without increasing alert burden or inequity.



\IEEEtriggeratref{18}
\bibliographystyle{IEEEtran}
\bibliography{references}

\end{document}